\documentclass[11pt]{article}

\newcommand{\be}{\begin{equation}}
\newcommand{\ee}{\end{equation}}
\newcommand{\bea}{\begin{eqnarray}}
\newcommand{\eea}{\end{eqnarray}}
\newcommand{\ba}{\begin{array}}
\newcommand{\ea}{\end{array}}
\newcommand{\hone}{h_{1,1}}
\newcommand{\htwo}{h_{2,1}}
\newcommand{\M}{\mathcal{M}}
\newcommand{\N}{\mathcal{N}}
\newcommand{\normomega}{\Vert\Omega\Vert}
\newcommand{\vcy}{V_{CY}}

\newcommand{\R}{{\sf R\hspace*{-0.9ex}%
  \rule{0.15ex}{1.5ex}\hspace*{0.9ex}}}

\begin{document}

\thispagestyle{empty}
\vspace{40pt}
\hfill{hep-th/0502112}

\vspace{128pt}

\begin{center}
\textbf{\Large Five dimensional 2-branes from special \\ \vspace{10pt} Lagrangian wrapped M5-branes }\\

\vspace{40pt}

Moataz H. Emam\footnote{Electronic address: memam@mtholyoke.edu}

\vspace{12pt} \textit{Department of Physics}\\
\textit{Mount Holyoke College}\\
\textit{South Hadley, MA 01075}\\
\end{center}

\vspace{40pt}

\begin{abstract}

We present an ansatz for black 2-branes in five dimensional
$\mathcal{N}=2$ supergravity theory that carry magnetic charge
with respect to general hypermultiplet scalars. We find explicit
solutions in certain special cases and examine the constraints on
the general case. These branes may be thought of as arising from
M-branes by wrapping eleven dimensional supergravity over special
Lagrangian calibrated cycles of a Calabi-Yau $3$-fold.

\end{abstract}

\newpage


\section{Introduction}

In this paper, we study black $2$-branes in the context of
ungauged $D=5$ $\mathcal{N}=2$ supergravity theory that arise from
dimensionally reducing $D=11$ supergravity on a Calabi-Yau
$3$-fold $\mathcal{M}$ \cite{Cadavid:1995bk}.  These branes carry
magnetic charge with respect to the hypermultiplet scalars and are
dual to the electrically charged instantons discussed in
\cite{Gutperle:2000sb,Gutperle:2000ve}. From a higher dimensional
viewpoint they correspond to M-branes wrapping special Lagrangian
(SLAG) cycles of $\mathcal{M}$.

In recent years, interest in nonperturbative solutions to
$\mathcal{N}=2$ supersymmetric theories has increased due to their
relevance to the conjectured equivalence between string theory on
anti-de Sitter space and certain superconformal gauge theories
living on the boundary of the space (the AdS/CFT duality)
\cite{Maldacena:2001uc}. In $\mathcal{N}=2$ supergravity (SUGRA),
however, an abundance of studies of solitonic solutions coupled to
the vector multiplets exist (for example \cite{Sabra:1997yd,
Sabra:1997dh} and other sources cited below), while very little
work on the hypermultiplets sector has been produced (for example
\cite{Gutperle:2000ve}). It then becomes of particular importance
to fill in this gap, and this paper presents a step in this
direction.

Another motivation for this work is to extend recent analysis of
supergravity solutions for branes wrapping K\"{a}hler calibrated
cycles to the SLAG case: $D=11$ Bogomol'nyi-Prasad-Sommerfield
(BPS) spacetimes, known as Fayyazuddin-Smith (FS) spacetimes
\cite{Fayyazuddin:1999zu}, that describe M2 and M5-branes wrapping
K\"{a}hler calibrated cycles of Calabi-Yau manifolds have been
discussed in
\cite{Fayyazuddin:1999zu,Gomberoff:1999ps,Fayyazuddin:2000em,Brinne:2000fh,Brinne:2000nf,Cho:2000hg,
Husain:2002tk,Husain:2003df,Husain:2003ag}. FS spacetimes
\cite{Fayyazuddin:1999zu} are specified entirely in terms of a
Hermitian metric $g_{m\bar n}$ on the Calabi-Yau space
$\mathcal{M}$, which varies as a function of the non-compact
coordinates transverse to the wrapped brane. Far from the brane in
the transverse space, the Hermitian metric $g_{m\bar n}$
asymptotes to a fixed Ricci-flat K\"{a}hler metric describing the
Calabi-Yau vacuum.  Moving closer to the brane, variation of
$g_{m\bar n}$ describes how the wrapped brane distorts the shape
of the Calabi-Yau space. Near the brane, the metric $g_{m\bar n}$
at a fixed transverse position is not even approximately
Ricci-flat (see, for example, \cite{Cho:2000hg}).

Supersymmetry requires that the Hermitian metric $g_{m\bar n}$
satisfy certain additional properties that depend on the precise
case under consideration.  For example, for M2-branes wrapping
$2$-cycles of Calabi-Yau $N$-folds with $N=2,3,4,5$, it was shown
in \cite{Husain:2002tk} that the K\"{a}hler form $J=i g_{m\bar n}
dz^m\wedge dz^{\bar n}$ at a fixed transverse position satisfies
the condition
\begin{equation}\label{kahler}
dJ^{N-1}=0.
\end{equation}
The metric also satisfies a nonlinear equation of motion, the
precise form of which again depends on the dimensions of the
Calabi-Yau and on the dimensions of the brane and the wrapped
cycles.  Explicit solutions to the FS equations of motion are
difficult to find\footnote{Exact solutions have been found only in
near horizon limits \cite{Fayyazuddin:2000em, Maldacena:2000mw,
Gauntlett:2000ng, Acharya:2000mu, Gauntlett:2001qs,
Gauntlett:2001jj}.}. On the other hand, one can gain intuition by
studying approximate solutions that hold far away from the wrapped
brane and correspond to black brane solutions of the dimensionally
reduced theory. If we focus on Calabi-Yau 3-folds, for example,
then M2-branes wrapping 2-cycles are black holes of the
corresponding $D=5$ $\mathcal{N}=2$ supergravity theory, and
M5-branes wrapping 4-cycles are black strings
\cite{Kastor:2003jy}. Such black hole and black string spacetimes
have been studied
\cite{Sabra:1997yd,Sabra:1997dh,Behrndt:1997ny,Sabra:1997kq,Behrndt:1997fq}
in the context of the attractor phenomenon of $D=4,5$
$\mathcal{N}=2$ supergravity discovered in
\cite{Ferrara:1995ih,Strominger:1996kf,Ferrara:1996dd,Ferrara:1996um}.

Recently, an ansatz for $D=11$ FS spacetimes corresponding to
M5-branes wrapping SLAG cycles of a Calabi-Yau $3$-fold has been
constructed \cite{Martelli:2003ki} using the G-structures approach
\cite{Gauntlett:2002sc,Gauntlett:2002nw, Gurrieri:2002wz,
Cardoso:2002hd,Gauntlett:2002fz}.  The Calabi-Yau metric in this
case also satisfies a nonlinear equation that is difficult to
solve exactly.  The black 2-branes introduced here are the
dimensionally reduced counterparts of these $D=11$ spacetimes. In
future work we plan to analyze the relation between these results.

The paper is structured as follows: Section two begins with a
review of $D=5$ $\mathcal{N}=2$ SUGRA theory coupled to the
universal hypermultiplet. We then proceed to discuss our solutions
(section 2.2) representing $D=5$ black 2-branes coupled to the
universal hypermultiplet. There are two such special cases,
corresponding to the dimensional reduction of M2 and M5 branes
over a six dimensional torus. Section 3 considers the theory with
the full hypermultiplets spectrum. For this, we review the special
geometry that arises in the space of moduli of the complex
structures of the Calabi-Yau space, following the notation of
\cite{Gutperle:2000ve}. We construct the form of the general
ansatz in section 4, representing a two brane coupled to the full
set of hypermultiplets, and find the differential equations
constraining the solution. These depend on the explicit form of
the underlying compact Calabi-Yau space. Finally, we conclude and
motivate further investigation.

\section{\label{univhyper}The universal hypermultiplet solutions}

Dimensionally reducing $D=11$ supergravity on a Calabi-Yau
$3$-fold $\M$ yields $D=5$ $\mathcal{N}=2$ supergravity coupled to
$(\hone-1)$ vector multiplets and $(\htwo+1)$ hypermultiplets
\cite{Cadavid:1995bk}; the $h$'s being the Hodge numbers of $\M$.
M-branes wrapping K\"{a}hler calibrated cycles of $\M$ deform the
K\"{a}hler structure of $\M$ and reduce to configurations in which
the vector multiplets are excited. SLAG wrapped M-branes, on the
other hand, our focus in this paper, deform the complex structure
of $\M$ and reduce to configurations carrying charge under the
hypermultiplet scalars. The two sectors of the theory decouple and
we only keep the hypermultiplets in our presentation. The
universal hypermultiplet is present even if $\htwo$ vanishes, and
in this section we present $D=5$ 2-branes for which only the
scalar fields of the universal hypermultiplet are excited. We will
see that these solutions are easily obtained by compactifying
known M-brane solutions of $D=11$ supergravity on a 6-torus.

We will use the following notation: The volume form is related to
the totally antisymmetric Levi-Civita symbol by $\varepsilon _{\mu
_1 \mu _2 \cdots \mu _D}  = e\bar \varepsilon _{\mu
    _1 \mu _2  \cdots \mu _D}$, where $e = \sqrt {\det \left(g_{\mu \nu }\right) }$ and $\bar \varepsilon _{012
\cdots }  =  + 1$.

\subsection{$D=5$ ${\mathcal N}=2$ supergravity coupled to the universal hypermultiplet}

We briefly sketch the derivation of the $D=5$ Lagrangian following
the discussion in \cite{Gutperle:2000sb}. The bosonic fields of
$D=11$ supergravity theory are the metric and the $3$-form gauge
potential $A$. The action is given by
\begin{equation}\label{eleven}
    S_{11}  = \frac{1}{{2\kappa _{11}^2 }}\int {d^{11} x\sqrt { - G}
    \left( {R - \frac{1}{{48}}F^2 } \right)}  - \frac{1}{12\kappa _{11}^2 } \int {A \wedge F \wedge
    F}.
\end{equation}
The $D=5$ couplings of gravity to the universal hypermultiplet follow from $D=11
$ fields of the form
\bea\label{universal-metric}
ds^2 & = & e^{-\sigma/3}ds_{CY}^2+ e^{2\sigma/3}g_{\mu\nu}dx^\mu dx^\nu,\quad\quad\mu,\nu=0,\ldots,4 \nonumber\\
A & = & {1\over 3!}A_{\mu\nu\rho} dx^\mu dx^\nu dx^\rho + {1\over
\sqrt{2} \normomega}\chi\Omega + {1\over \sqrt{2}
\normomega}\bar\chi\bar\Omega, \eea
where $ds_{CY}^2$ is a fixed Ricci flat metric on the Calabi-Yau
space $\M$ and $\Omega$ is the holomorphic $3$-form on $\M$ with
norm $\normomega$. The $D=5$ fields of the universal
hypermultiplet are the real overall volume modulus $\sigma$ of the
the Calabi-Yau space, the $D=5$ 3-form gauge potential
$A_{\mu\nu\rho}$ and the complex scalar $\chi$.  The $D=5$ action
is given by ($\kappa_5^2=\kappa_{11}^2/\vcy$):
\begin{equation}\label{universal}
    S_5={1\over 2\kappa_5^2} \int d^5x\sqrt{-g} \left[ R-{1\over
2}(\nabla\sigma)^2 -{1\over 48} e^{-2\sigma} F^2 -  e^\sigma
 |\nabla\chi|^2\right]
-{1\over 4\kappa_5^2}\int F\wedge(\chi d\bar\chi-\bar\chi d\chi).
\end{equation}

The equations of motion of $\sigma$, $F_{\mu \nu \rho \sigma}$ and
$(\chi, \bar \chi)$ are, respectively:
\begin{eqnarray}
    \nabla ^2 \sigma  -  \frac{1}{2}e^\sigma  \left( {\partial _\mu  \chi } \right)\left(
    {\partial     ^\mu  \bar \chi } \right)+ \frac{1}{{24}}e^{ - 2\sigma } F_{\mu \nu
    \rho     \sigma } F^{\mu \nu \rho \sigma }  &=& 0\nonumber \\
    \nabla ^\mu  \left( {e^{ - 2\sigma } F_{\mu \nu \rho \sigma}  +
    \bar\varepsilon_{\mu \nu \rho \sigma \alpha} \frac{i}{2}\left[ {\chi \left(
    {\partial^\alpha  \bar \chi } \right) - \bar \chi \left( {\partial^\alpha  \chi
    }     \right)} \right]  } \right) &=& 0
    \nonumber \\
    \nabla ^\mu  \left[ {e^\sigma  \left( {\partial _\mu   \chi }
    \right) + \frac{i}{{48}}\bar \varepsilon _{\mu \nu \rho \sigma
    \alpha } F^{\nu \rho \sigma \alpha }  \chi } \right] &=& -\frac{i}{{48}}\bar \varepsilon _{\mu \nu \rho \sigma \alpha }
    F^{\mu \nu \rho \sigma } \left( {\partial ^\alpha   \chi }
    \right) \nonumber \\
    \nabla ^\mu  \left[ {e^\sigma  \left( {\partial _\mu  \bar \chi }
    \right) - \frac{i}{{48}}\bar \varepsilon _{\mu \nu \rho \sigma
    \alpha } F^{\nu \rho \sigma \alpha } \bar \chi } \right] &=& +\frac{i}{{48}}\bar \varepsilon _{\mu \nu \rho \sigma \alpha }
    F^{\mu \nu \rho \sigma } \left( {\partial ^\alpha  \bar \chi }
    \right). \nonumber \\ \label{chieom}
\end{eqnarray}

The full action is invariant under the set of supersymmetry (SUSY)
transformations:
\begin{eqnarray}
    \delta \psi _\mu ^1  &=&  \left( {\partial _\mu \epsilon_1 } \right) + \
    \frac{1}{4}\omega     _\mu^{\;\;\;\hat \mu\hat \nu} \Gamma _{\hat \mu\hat \nu}
    \epsilon_1  + i \frac{e^{-\sigma}}{{96}} \varepsilon _{\mu
    \nu  \rho \sigma \lambda } F^{\nu \rho \sigma \lambda }  \epsilon
    _1   - \frac{e^{\frac{\sigma }{2}}}{{\sqrt 2 }}\left( {\partial _\mu  \chi } \right)\epsilon _2 \nonumber \\
    \delta \psi _\mu ^2  &=&  \left( {\partial _\mu \epsilon_2 } \right) + \
    \frac{1}{4}\omega     _\mu^{\;\;\;\hat \mu\hat \nu} \Gamma _{\hat \mu\hat \nu}
    \epsilon_2  - i\frac{e^{-\sigma}}{{96}} \varepsilon _{\mu
    \nu  \rho \sigma \lambda } F^{\nu \rho \sigma \lambda }  \epsilon
    _2 + \frac{e^{\frac{\sigma }{2}}}{{\sqrt 2 }}\left( {\partial _\mu  \bar\chi } \right)\epsilon _1  \nonumber \\
    \delta \xi _1  &=& \frac{1}{2}\left( {\partial _\mu  \sigma } \right)\Gamma
    ^\mu      \epsilon _1 - \frac{i}{{48}}e^{ - \sigma } \varepsilon _{\mu \nu \rho \sigma \lambda
    }     F^{\mu \nu \rho \sigma } \Gamma ^\lambda  \epsilon _1  +
    \frac{e^{\frac{\sigma }{2}}}{{\sqrt 2 }} \left( {\partial _\mu  \chi } \right)\Gamma
    ^\mu  \epsilon _2  \nonumber \\
    \delta \xi _2  &=& \frac{1}{2}\left( {\partial _\mu  \sigma } \right)\Gamma
    ^\mu      \epsilon _2 + \frac{i}{{48}}e^{ - \sigma } \varepsilon _{\mu \nu \rho \sigma \lambda
    }     F^{\mu \nu \rho \sigma } \Gamma ^\lambda  \epsilon _2  -
    \frac{e^{\frac{\sigma }{2}}}{{\sqrt 2 }} \left( {\partial _\mu  \bar\chi } \right)\Gamma
    ^\mu  \epsilon _1, \label{hyperSUSY}
\end{eqnarray}
where $\psi$ and $\xi$ are the gravitini and hyperini fermions
respectively, the $\epsilon$'s are the $\mathcal{N}=2$ SUSY
spinors, $\omega$ is the spin connection and the hatted indices
are frame indices in a flat tangent space.

\subsection{\label{univhypersolution}The solutions}

We find explicit solutions with only the universal hypermultiplet
fields excited. The first one may be thought of as a direct
reduction of the M2-brane ({\it i.e.} without wrapping) to $D=5$.
The $D=11$ M2-brane solution is given by
\begin{eqnarray}\label{mtwo}
ds_{11}^2 &=& f^{-2/3}(-dx_0^2+dx_1^2+dx_2^2) + f^{1/3} ( {dx_3^2
+  \cdots  + dx_{10}^2 } )
\nonumber\\
A_{012}&=&f^{-1},
\end{eqnarray}
where for the direct reduction to $D=5$ we take $f=f(x_3,x_4)$
with $(\partial_3 ^2+\partial_4^2)f=0$. If we let
$x^5,\dots,x^{10}$ be coordinates on a flat $6$-torus, then
putting this into the form of the ansatz for the universal
hypermultiplet we find
\begin{eqnarray}\label{mtworeduced}
ds^2_5 &=& -dx_0^2+dx_1^2+dx_2^2 + f(dx_3^2+dx_4^2) \nonumber\\
e^\sigma &=& f^{-1},\qquad A_{012}= f^{-1},\qquad \chi=\bar\chi=0.
\end{eqnarray}
This satisfies both the $D=5$ equations of motion (\ref{chieom})
and the SUSY equations (\ref{hyperSUSY}) in a straightforward way,
as can be easily checked.

Another more interesting solution we find is one which may be
thought of as the wrapping of a M5-brane on a particular SLAG
cycle of the 6-torus. Exciting only the universal hypermultiplet
requires a configuration of 4 sets of intersecting M5-branes as
follows
\be\label{intersect}
\matrix{
M5:&1&2& & &5& 6& 7 & & & \cr
\overline{M5}:&1 &2 & & &5& & & &9&10\cr
M5:&1&2 & & & &6 & &8& & 10\cr
\overline{M5}:&1&2 & & & &  &7 &8&9 &  \cr}
\ee
The $D=11$ metric and $6$-form gauge potential can then be
constructed as follows
\begin{eqnarray}\label{mfive}
ds_{11}^2 &=& h^{-4/3}(-dx_0^2+dx_1^2+dx_2^2) + h^{8/3}(dx^2_3 +dx_4^2)
+h^{2/3}(dx_5^2+\dots +dx_{10}^2)\nonumber\\
A&=&{1\over 2}h^{-1} dx^0\wedge dx^1\wedge dx^2\wedge (\Omega +\bar\Omega)
\end{eqnarray}
where $\Omega= dz^1\wedge dz^2\wedge dz^3$ is the holomorphic
$3$-form associated with the complex coordinates $z^1=x^5+ix^8$,
$z^2=x^6+ix^9$ and $z^3=x^7+ix^{10}$ on the $6$-torus. This
satisfies the $D=11$ equations of motion and SUSY variation
equations provided that the supersymmetry parameters satisfy
$\epsilon = h^{-1/3} \epsilon_0$, where $\epsilon_0$ is a constant
spinor.

The corresponding $D=5$ fields are then found to be
\begin{eqnarray}\label{mfivereduced}
ds^2_5 &=& -dx_0^2+dx_1^2+dx_2^2 + h^4(dx_3^2+dx_4^2) \nonumber\\
e^\sigma &=& h^{-2}, \qquad A_{\mu\nu\rho}= 0,\qquad
(\partial_\mu\chi)=i\sqrt{2}\;\bar\varepsilon_\mu^{\;\
\nu}(\partial_\nu h),\qquad
(\partial_\mu\bar\chi)=-i\sqrt{2}\;\bar\varepsilon_\mu^{\;\
\nu}(\partial_\nu h),
\end{eqnarray}
where the equations of motion are satisfied provided that the
function $h$ is harmonic in the transverse space. The SUSY
equations are also satisfied provided the constant spinors are
simply related by $\epsilon_1=\pm i \epsilon_2$.

\section{\label{specialgeometry}Special geometry of the complex structure moduli
 space}

The $D=5$ $\mathcal{N}=2$ supergravity Langrangian including the
full set of $(\htwo+1)$ hypermultiplets can be written in terms of
geometric quantities on the moduli space of the complex structures
on the Calabi-Yau manifold $\M$.  These structures are discussed
in detail in \cite{Candelas:1990pi} and we will give a brief
review here. Start by taking a basis of the homology $3$-cycles
$(A^I,B_J)$ with $I,J=0,1,\dots,\htwo$ and a dual cohomology basis
of $3$-forms $(\alpha_I,\beta^J)$ such that
\begin{equation}
\int_{A^J}\alpha_I=\int_{\M}\alpha_I\wedge\beta^J=\delta_I^J,\qquad
\int_{B_I}\beta^J=\int_{\M}\beta^J\wedge\alpha_I=-\delta_I^J.
\end{equation}
Define the periods of the holomorphic $3$-form $\Omega$ on $\M$ by
\begin{equation}
Z^I=\int_{A^I}\Omega,\qquad   F_I=\int_{B_I}\Omega.
\end{equation}
The periods $Z^I$ can be regarded as coordinates on the complex
structure moduli space.  Since $\Omega$ can be multiplied by an
arbitrary complex number without changing the complex structure,
the $Z^I$ are projective coordinates.  The remaining periods $F_I$
can then be regarded as functions $F_I(Z)$.  One can further show
that $F_I$ is the gradient of a function $F(Z)$, known as the
prepotential, that is homogeneous of degree two in the
coordinates, {\it i.e.} $F_I=\partial_I F(Z)$ with $F(\lambda
Z)=\lambda^2 F(Z)$. The quantity $F_{IJ}(Z)=\partial_I\partial_J
F(Z)$ will also play an important role. Non-projective coordinates
can then be given by taking {\it e.g.} $z^i=Z^i/Z^0$ with
$i=1,\dots,\htwo$. The K\"{a}hler potential of the complex
structure moduli space is $\mathcal{K}=-\log
(i\int_\M\Omega\wedge\bar\Omega)$. Given the expansion of $\Omega$
in terms of the periods
\begin{equation}
\Omega=Z^I\alpha_I-F_I\beta^I,
\end{equation}
the K\"{a}hler potential is determined in terms of the
prepotential $F(Z)$ according to
\begin{equation}
\mathcal{K} = -\ln\left[ i ( Z^I \bar F_I - \bar Z^I F_I )\right].
\end{equation}

An important ingredient in carrying out the dimensional reduction
is the expression for the Hodge-K\"{a}hler duals of the cohomology
basis $(\alpha_I,\beta^J)$ expressed in terms of the basis itself.
This is given by
\begin{eqnarray}
{}^*\alpha_I & = & \left( \gamma_{IJ} +
\theta_{IK}\gamma^{KL}\theta_{LJ}\right)\beta^J-\theta_{IK}\gamma^{KJ}\alpha_J
 \nonumber\\
{}^*\beta^I & = &
\gamma^{IK}\theta_{KJ}\beta^J-\gamma^{IJ}\alpha_J.\label{cohomologyduals}
\end{eqnarray}
Here $\theta_{IJ}$ and $\gamma_{IJ}$ are real matrices defined by
\begin{eqnarray}
\N_{IJ} &=& \bar F_{IJ} -2 i {N_{IK}Z^KN_{JL}Z^L\over Z^P N_{PQ} Z^Q}\nonumber\\
   & =& \theta_{IJ}-i \gamma_{IJ}
\end{eqnarray}
where $N_{IJ}=Im(F_{IJ})$, $\gamma^{IJ}\gamma_{JK}=\delta^I_K$ and
$\N_{IJ}$ is known as the period matrix.

\section{The general case}
The derivation of the Lagrangian for the bosonic fields of the
$D=5$ theory is sketched in \cite{Gutperle:2000ve}, where
Euclidean instantons carrying electric charge with respect to the
hypermultiplet scalars are constructed. We follow the notation of
\cite{Gutperle:2000ve}, with the exception that we will be working
in Lorentzian signature. The reduction of the $D=11$ action is
done over the metric (\ref{universal-metric}), and the $D=11$
3-form is now expanded in terms of the cohomology basis as follows
\begin{equation}
    A = {1\over 3!}A_{\mu\nu\rho} dx^\mu dx^\nu dx^\rho + \sqrt 2 \left( {\zeta
^I \alpha _I  + \tilde \zeta _I \beta ^I }
    \right), \qquad \mu,\nu=0,\ldots,4.
\end{equation}

The resulting $D=5$ action for the bosonic fields is
\begin{eqnarray}
    S_5  &=& \frac{1}{{2\kappa _{5}^2 }}\int {d^5 x\sqrt { - g} \left[ { R - \frac{1}{2}\left( {\partial
    _\mu \sigma     } \right)\left( {\partial ^\mu  \sigma } \right) - G_{i\bar j} (
    {\partial _\mu  z^i } )( {\partial ^\mu  z^{\bar j} } )}
    \right.}  \nonumber \\
    &-& \left.{ \frac{1}{{48}}e^{ - 2\sigma } F_{\mu \nu \rho \sigma } F^{\mu \nu \rho \sigma }-\frac{1}{{24}}\bar\varepsilon _{\mu \nu \rho \sigma \alpha } F^{\mu \nu
    \rho         \sigma } K^\alpha  \left( {\zeta ,\tilde \zeta } \right) + e^\sigma
    L_\mu     ^\mu  \left( {\zeta ,\tilde \zeta } \right)} \right], \label{d5theory}
\end{eqnarray}
where we have defined:
\begin{eqnarray}
    K_\alpha  ( {\zeta ,\tilde \zeta } ) &=& \left[ {\zeta ^I (
    {\partial_\alpha  \tilde \zeta _I } ) - \tilde \zeta _I ( {\partial_\alpha  \zeta
    ^I     } )} \right] \nonumber \\
    L_{\mu \nu}  ( {\zeta ,\tilde \zeta } ) &=&  - \left( {\gamma  + \gamma ^{
    -     1} \theta ^2 } \right)\left(
    {\partial _\mu  \zeta } \right)\left( {\partial _\nu  \zeta } \right)
    - \gamma ^{ - 1}( {\partial _\mu  \tilde \zeta } )( {\partial _\nu
    \tilde \zeta     } ) \nonumber \\
    & & -2\gamma ^{ - 1} \theta \left( {\partial _\mu  \zeta } \right)(
    {\partial _\nu  \tilde \zeta } ). \label{KandL}
\end{eqnarray}

The scalar fields $z^i$, $z^ {\bar i}$ with $i=1,\dots,\htwo$ are
complex coordinates on the complex structure moduli space with
metric $G_{i\bar j}$. The scalar fields $\zeta^I,\tilde\zeta_I$
with $I=0,1,\dots,\htwo$ arise from the dimensional reduction of
the $D=11$ $3$-form gauge potential.  The scalar field $\sigma$ is
the overall volume scalar of the Calabi-Yau $\M$ and $F_{\mu \nu
\rho \sigma }$ is the $D=5$ 4-form field strength. Each
hypermultiplet has 4 scalar fields.  The scalar fields $(z^i,z^
{\bar i},\zeta^i,\tilde \zeta_i)$ make up $\htwo$ of the
hypermultiplets.  The additional universal hypermultiplet is
comprised of the fields $(a,\sigma,\zeta^0,\tilde \zeta_0)$, where
the axion $a$ is the scalar dual of the 3-form gauge potential
$A_{\mu\nu\rho}$.

The equations of motion for $\sigma$, $F_{\mu \nu \rho \lambda}$,
$(z,\bar z )$ and $( {\zeta ,\tilde \zeta } )$ are:
\begin{eqnarray}
    \nabla ^2 \sigma  + e^\sigma  L_\mu ^\mu + \frac{1}{{24}}e^{ - 2\sigma } F_{\mu
    \nu     \rho \sigma }   F^{\mu     \nu \rho \sigma } &=& 0 \nonumber \\
    \nabla ^\mu  \left( {e^{ - 2\sigma }  F_{\mu \rho \sigma \lambda}  +
    \bar \varepsilon _{\mu \rho \sigma \lambda \nu} K^\nu  } \right) &=& 0
    \nonumber     \\
    \nabla ^2 z^i  + \Gamma _{jk}^i \left( {\partial _\alpha  z^j } \right)\left(
    {\partial     ^\alpha  z^k } \right) + e^\sigma  \left( {\partial ^i L_\mu ^\mu  }
    \right) &=& 0 \nonumber \\
    \nabla ^2 z^{\bar i}  + \Gamma _{\bar j\bar k}^{\bar i} ( {\partial _\alpha
    z^{\bar j} } )( {\partial ^\alpha  z^{\bar k} } ) + e^\sigma
    (     {\partial ^{\bar i} L_\mu ^\mu  } ) &=& 0
    \nonumber \\
    \nabla ^\mu  \left[ {e^\sigma  \left( {\gamma  + \gamma ^{ - 1}
    \theta ^2 } \right)\left( {\partial _\mu  \zeta } \right) + \gamma
    ^{ - 1} \theta e^\sigma  ( {\partial _\mu  \tilde \zeta }
    )}\right. &-& \left.{ \frac{{ 1 }}{{48}}\bar \varepsilon _{\mu \nu \rho \sigma
    \alpha } F^{\mu \nu \rho \sigma } \tilde \zeta } \right] \nonumber\\&=& \frac{1}{{48}}\bar \varepsilon _{\mu\nu \rho \sigma \alpha
    } F^{\mu \nu \rho \sigma } ( {\partial ^\alpha \tilde
    \zeta } ) \nonumber \\
    \nabla ^\mu  \left[ {e^\sigma  \gamma ^{ - 1} \theta \left(
    {\partial _\mu  \zeta } \right) + e^\sigma  \gamma ^{ - 1} (
    {\partial _\mu  \tilde \zeta } )}\right. &+& \left.{ \frac{{e^\sigma
    }}{{48}}\bar \varepsilon _{\mu \nu \rho \sigma \alpha } F^{\mu \nu
    \rho \sigma } \zeta } \right] \nonumber\\&=&  - \frac{1}{{48}}
    \bar \varepsilon _{\mu \nu \rho \sigma \alpha } F^{\mu \nu \rho
    \sigma} \left( {\partial ^\alpha  \zeta } \right).
    \label{xieom}
\end{eqnarray}

Further study of the structure of the theory (see
\cite{Gutperle:2000ve} and the references within) reveals that the
hypermultiplets define a $(\htwo+1)$ dimensional quaternionic
space. This structure, in five dimensions, is dual to the special
K\"{a}hler geometry of the $D=4$ vector multiplets sector via the
so called c-map ({\it e.g.} \cite{DeJaegher:1997ka}). This duality
justifies the use of the special geometry defined in
\S\ref{specialgeometry} as opposed to the explicit quaternionic
form.

Furthermore, one finds that the theory is invariant under the
symplectic group $Sp(2\htwo,\R)$, {\it i.e.} (\ref{d5theory})
actually defines a family of Lagrangians that differ from each
other only by a rotation in symplectic space that has no effect on
the physics. In fact, if we define
\begin{equation}\label{symvec}
    V = \left( {\begin{array}{*{20}c}
       {L^I }  \\
       {M_J }  \\
    \end{array}} \right) \equiv e^{{\mathcal{K} \mathord{\left/
     {\vphantom {\mathcal{K} 2}} \right.
     \kern-\nulldelimiterspace} 2}} \left( {\begin{array}{*{20}c}
       {Z^I }  \\
       {F_J }  \\
    \end{array}} \right)
\end{equation}
satisfying
\begin{equation}\label{covderiv}
    \nabla _{\bar i} V = \left[ {\partial _{\bar i}  -
    \frac{1}{2}\left( {\partial _{\bar i} \mathcal{K}} \right)} \right]V =
    0,
\end{equation}
then $V$ is a basis vector in symplectic space that satisfies the
inner product
\begin{equation}\label{innerprod}
    i\left\langle V \right|\left. {\bar V} \right\rangle  = i\left(
    {\bar L^I M_I  - L^I \bar M_I } \right) = 1.
\end{equation}
An orthogonal vector may be defined by
\begin{equation}\label{covderiv1}
    U_i  \equiv \left(
    {\begin{array}{*{20}c}
       {f_i^I }  \\
       {h_{J|i} }  \\
    \end{array}} \right)=\nabla _i V,
\end{equation}
such that
\begin{equation}
    \left\langle V \right|\left. {U_i } \right\rangle  = \left\langle V \right|\left. {U_{\bar i} } \right\rangle  = 0.
\end{equation}

Based on this, the following useful identities may be derived:
\begin{eqnarray}
    \mathcal{N}_{IJ} L^J  &=& M_I \quad ,\quad \mathcal{N}_{IJ} f_i^J  =
    h_{I|i}\nonumber\\
    \gamma ^{IJ}  &=& 2\left( {G^{i\bar j} f_i^I f_{\bar j}^J  + L^I \bar L^J }
    \right) \nonumber\\
    \left( {\nabla _{\bar j} f_i^I } \right) &=& G_{i\bar j} L^I ,\quad \left( {
\nabla _{\bar j} h_{iI} } \right) = G_{i\bar j} M_I
    \nonumber \\
    \gamma _{IJ} L^I \bar L^J  &=&   \frac{1}{2} \nonumber \\
    G_{i\bar j}  &=&    2f_i^I \gamma _{IJ} f_{\bar
    j}^J.\label{computational}
\end{eqnarray}

The gravitini variation equations are:
\begin{eqnarray}
    \delta \psi _\mu ^A  &=& ( {\partial _\mu \epsilon^A } ) + \
    \frac{1}{4}\omega     _\mu^{\;\;\;\hat \mu\hat \nu} \Gamma _{\hat \mu\hat \nu}
    \epsilon^A + \left[ {\mathcal{Q}_\mu  } \right]_{\;\;B}^A \epsilon ^B  \nonumber\\
    \left[ {\mathcal{Q}_\mu  } \right] &=& \left[ {\begin{array}{*{20}c}
    {\frac{1}{4}\left( {v_\mu   - \bar v_\mu   - \frac{{\bar Z^I N_{IJ} \left( {\partial _\mu  Z^J } \right) -
    Z^I N_{IJ} \left( {\partial _\mu  \bar Z^J } \right)}}{{\bar Z^I N_{IJ} Z^J
    }}  } \right)} & { - \bar
    u_\mu     }  \\
    {u_\mu     } & { - \frac{1}{4}\left( {v_\mu   - \bar v_\mu   - \frac{{\bar Z^I N_{IJ} \left( {\partial _\mu  Z^J } \right) -
    Z^I N_{IJ} \left( {\partial _\mu  \bar Z^J } \right)}}{{\bar Z^I N_{IJ} Z^J
    }}  } \right)}
    \\
    \end{array}} \right] \nonumber \\ \label{gravitinotrans}
\end{eqnarray}
where the indices $A$ and $B$ run over $(1,2)$, and
\begin{eqnarray}
    u_\mu   &=&  - ie^{\frac{\sigma }{2}} \left[ {M_I ( {\partial _\mu
    \zeta^I } ) + L^I ( {\partial _\mu  \tilde \zeta _I } )}
    \right] \nonumber \\
    \bar u_\mu   &=& + ie^{\frac{\sigma }{2}} \left[ {\bar M_I ( {\partial
    _\mu \zeta ^I } ) + \bar L^I ( {\partial _\mu  \tilde \zeta _I }
    )} \right] \nonumber \\
    v_\mu   &=& \frac{1}{2}\left( {\partial _\mu  \sigma } \right)
    + \frac{i}{2}e^\sigma  \left[ {\left( {\partial _\mu  a} \right) - K_\mu  } \right]\nonumber \\
    \bar v_\mu   &=& \frac{1}{2}\left( {\partial _\mu  \sigma } \right) -
    \frac{i}{2}e^\sigma  \left[ {\left( {\partial _\mu  a} \right) - K_\mu  }
    \right].\label{eqns5}
\end{eqnarray}

The hyperini equations are:
\begin{eqnarray}
    \delta \xi _1^I  = e_{\;\;\mu} ^{1I} \Gamma ^\mu  \epsilon _1  - \bar e_{\;\;\mu}^{2I}
    \Gamma ^\mu  \epsilon _2  \nonumber \\
    \delta \xi _2^I  = e_{\;\;\mu} ^{2I} \Gamma ^\mu  \epsilon _1  + \bar e_{\;\;\mu}^{1I}
    \Gamma ^\mu  \epsilon _2, \label{hyperinotrans}
\end{eqnarray}
written in terms of the beins:
\begin{eqnarray}
    e_{\;\;\mu} ^{1I}  &=& \left( {\begin{array}{*{20}c}
   {u_\mu  }  \\
   {E_{\;\;\mu} ^{\hat i} }  \\
    \end{array}} \right) \nonumber \\\nonumber \\ e_{\;\;\mu} ^{2I}  &=& \left(
    {\begin{array}{*{20}c}
   {v_\mu  }  \\
   {e_{\;\;\mu} ^{\hat i} }  \\
    \end{array}} \right)
\end{eqnarray}
\begin{eqnarray}
    E_{\;\;\mu} ^{\hat i}  &=&  - ie^{\frac{\sigma }{2}} e^{\hat ij} \left[ {h_{jI} \left(
    {\partial _\mu  \zeta ^I } \right) + f_j^I \left( {\partial _\mu  \tilde \zeta _I }
    \right)} \right] \nonumber \\
    \bar E_{\;\;\mu} ^{\hat i}  &=&  + ie^{\frac{\sigma }{2}} e^{\hat i\bar j} \left[ {h_{\bar
    jI}     \left( {\partial _\mu  \zeta ^I } \right) + f_{\bar j}^I \left( {\partial
    _\mu      \tilde \zeta _I } \right)} \right],
\end{eqnarray}
and the beins of the special K\"{a}hler metric:
\begin{eqnarray}
    e_{\;\;\mu} ^{\hat i}  &=& e_{\;\;j}^{\hat i} \left( {\partial _\mu  z^j } \right)\quad,\quad \quad
    \quad \bar e_{\;\;\mu} ^{\hat i}  = e_{\;\;{\bar j}}^{\hat i} \left( {\partial _\mu  z^{\bar j} }
    \right) \nonumber \\
    G_{i\bar j}  &=& e_{\;\;i}^{\hat k} e_{\;\;{\bar j}}^{\hat l} \delta _{\hat k\hat l}.
\end{eqnarray}

\subsection{Constructing the ansatz}

Solitonic solutions coupled to the $\mathcal{N}=2$ hypermultiplets
are quite rare in the literature. On the other hand, there is an
abundance of solutions coupled to the vector multiplets. $D=4,5$
black holes coupled to vector multiplets, for example, have been
extensively studied. In what follows, we are only interested in
the case $A_{\mu\nu\rho}=0$ representing the wrapping of a
M5-brane.

For the scalar fields, we expect that their most general form will
be an expansion in terms of more than one harmonic function. In
the previous simpler cases, a single function, harmonic in the
transverse space, sufficed. We now need to generalize this by
introducing a number $2\left({\htwo+1}\right)$ electric charges
$q_I$ and a similar number of magnetic charges $\tilde q^I$,
defined by harmonic functions corresponding to each homology cycle
on the Calabi-Yau submanifold as follows:
\begin{equation}
    H_I  = h_I  + q_I \ln r\quad ,\quad \tilde H^I  = \tilde h^I  + \tilde q^I \ln
    r \quad ,\quad I = 0, \ldots ,h_{2,1},
\end{equation}
where $h$ and $\tilde h$ are constants and $r$ is the radial
coordinate in the two dimensional space transverse to the brane.
We find that the BPS 2-brane metric is given by
\begin{equation}
    ds^2  = \eta_{ab} dx^a dx^b  + e^{-2\sigma } \delta _{\mu \nu } dx^\mu   dx^\nu,\quad \quad a,b=0,1,2\quad \mu,\nu=3,4,
\end{equation}
satisfying both the SUSY and field equations, yielding:
\begin{eqnarray}
    \left( {\partial _\mu  \sigma } \right) &=&  - 2e^{ \frac{\sigma}{2} } \left
[
    {L^I     \left( {\partial _\mu  H_I } \right) - M_I \left( {\partial _\mu  \tilde H^I
    }     \right)} \right] \nonumber \\
    \left( {\partial _\mu  z^i } \right) &=& - e^{ \frac{\sigma}{2} } G^{i\bar j
}
    \left[     {f_{\bar j}^I \left( {\partial _\nu  H_I } \right) - h_{\bar jI}
\left(
    {\partial _\nu  \tilde H^I } \right)} \right] \nonumber \\
    \left( {\partial _\mu  z^{\bar i} } \right) &=& -e^{ \frac{\sigma}{2} } G^{\bar ij}    \left[ {f_j^I \left( {\partial _\mu  H_I } \right) - h_{jI} \left( {\partial _\mu
    \tilde H^I } \right)} \right]\nonumber\\
    \left( {\partial _\mu  \zeta ^I } \right) &=& \pm
    {\bar\varepsilon_\mu}^{\;\;\;\nu}   \left( {\partial
    _\nu      \tilde H^I } \right)\nonumber \\
    \left( {\partial _\mu  \tilde \zeta _I } \right) &=& \pm
    {\bar\varepsilon_\mu}^{\;\;\;\nu}  \left( {\partial _\nu  H_I } \right). \label{zetaeom}
\end{eqnarray}

The Bianchi identity on the $(\zeta, \tilde\zeta)$ fields gives
the harmonic condition on $(H, \tilde H)$, and the SUSY spinors
satisfy $\epsilon_1=\pm\epsilon_2$. One also finds that the
condition
\begin{equation}\label{grandsolution}
    H_I ( {\partial _\mu  \tilde H^I } ) - \tilde H^I
    \left( {\partial _\mu  H_I } \right) = h\tilde q - \tilde hq = 0
\end{equation}
is satisfied, guaranteeing the vanishing of the M5-brane charge.

We note that there is a relationship between the charges $q$ and
$\tilde q$ and the central charge $Z$ of the theory as follows
\cite{Witten:mh}:
\begin{eqnarray}
    Z &=& \left( {L^I q_I  - M_I \tilde q^I } \right) \nonumber \\
    \bar Z &=& \left( {\bar L^I q_I  - \bar M_I \tilde q^I }
    \right).
\end{eqnarray}

Based on this, equations (\ref{zetaeom}) become:
\begin{eqnarray}
    \frac{{d\sigma }}{{dr}} &=&  - 2e^{  \frac{\sigma}{2} } \frac{Z}{r}\nonumber \\
    \frac{{dz^i }}{{dr}} &=&  - e^{ -\frac{\sigma}{2} } \frac{{\nabla ^i \bar Z}}
{r}
    \nonumber \\
    \frac{{dz^{\bar i} }}{{dr}} &=&  - e^{  -\frac{\sigma}{2} } \frac{{\nabla^{\bar i}
    Z}}{r}.     \label{zzeom}
\end{eqnarray}

The solution may be further specified in terms of the moduli by
adopting Sabra's ansatz \cite{Sabra:1997dh}:
\begin{equation}
    H_I  = i\left( {F_I  - \bar F_I } \right)\quad \quad \tilde H^I =
    i( {Z^I  - \bar Z^I } )\quad \quad \sigma  =  - \mathcal{K}.
\end{equation}

The details of how this does indeed satisfy (\ref{zzeom}) are very
similar to those in Sabra's paper and will not be reproduced here.
Finally, one can easily check that these equations immediately
yield the universal hypermultiplet solution (\ref{mfivereduced})
by setting $\left({h_{2,1}=0}\right)$ and integrating
(\ref{zzeom}).

\section*{Conclusion}

We have discussed the coupling of BPS black 2-branes to the
hypermultiplets of five dimensional $\mathcal{N}=2$ supergravity
theory and found explicit solutions in the special case of the
universal hypermultiplet. In analyzing the general case, we have
used the fact that the quaternionic structure of the theory can be
mapped to the special geometry of the four dimensional vector
multiplets sector, allowing us to use the more familiar
symplectically invariant form of the action for the five
dimensional hypermultiplets, following the work of
\cite{Gutperle:2000ve}.

Those 2-branes can be seen as various types of M-branes wrapped
over supersymmetric cycles of Calabi-Yau manifolds. We have
particularly focused on the case of wrapped M5-branes. Further
analysis is left for future work.

\section*{Acknowledgments}
This paper is based on results from the author's doctoral
dissertation \cite{Emam:2004nc}, conducted under the patient
guidance of David Kastor at the University of Massachusetts
Amherst. I would like to extend him my thanks and gratitude for
his support and invaluable advice.

Special thanks are also due to Wafic Sabra of the American
University of Beirut, for useful discussions during the writing of
this paper and for taking the time to read the final draft,
providing me with further advice.

This work was supported in part by the National Science Foundation
grant NSF PHY0244801.

\end{document}